%% file: main.tex
\documentclass[conference]{IEEEtran}
\IEEEoverridecommandlockouts
\usepackage[noadjust]{cite}

\usepackage{amsmath,amssymb,amsfonts}
\usepackage{algorithmic}
\usepackage{graphicx}
\usepackage{textcomp}
\usepackage{xcolor}

\usepackage{url}
\usepackage{breakurl}

\usepackage{hyperref}

\hypersetup{
    colorlinks,
    linkcolor={[rgb]{0,0,0.4}},
    citecolor={[rgb]{0,0,0.4}},
    urlcolor={[rgb]{0,0,1}},
    breaklinks=true,
}

\usepackage[ruled,vlined,linesnumbered]{algorithm2e}
\usepackage{comment}
\usepackage{color}
\usepackage{graphicx}
\usepackage{caption}
\usepackage{subcaption}
\usepackage{threeparttable}
\usepackage{multirow}
\usepackage{booktabs}
\usepackage{verbatim}
\usepackage{epstopdf}
\usepackage{rotating}
\usepackage{listings}
\usepackage{listing}
\usepackage{paralist}
\usepackage{arydshln}

\usepackage{enumitem,amssymb}
\usepackage{balance}
\usepackage{endnotes}
\usepackage{xspace}
\usepackage{amsfonts}
\usepackage{pifont}
\usepackage{wasysym}
\usepackage{tikz}

\newcommand{\Bstrut}{\rule[-0.9ex]{0pt}{0pt}}   

\newcommand{\tick}{{\color{green!50!black}{\ding{51}}}}
\newcommand{\cross}{{\color{red}{\ding{55}}}}

\graphicspath{{./figs/}}

\def\BibTeX{{\rm B\kern-.05em{\sc i\kern-.025em b}\kern-.08em
    T\kern-.1667em\lower.7ex\hbox{E}\kern-.125emX}}
\begin{document}

\title{Scam Pandemic: \\How Attackers Exploit Public Fear through Phishing}

\author{\IEEEauthorblockN{Marzieh Bitaab\IEEEauthorrefmark{1},
Haehyun Cho\IEEEauthorrefmark{1}, Adam Oest\IEEEauthorrefmark{1}\IEEEauthorrefmark{2},
Penghui Zhang\IEEEauthorrefmark{1}, Zhibo Sun\IEEEauthorrefmark{1}, Rana Pourmohamad\IEEEauthorrefmark{1}, \\ Doowon Kim\IEEEauthorrefmark{3}, Tiffany Bao\IEEEauthorrefmark{1}, Ruoyu Wang\IEEEauthorrefmark{1}, Yan Shoshitaishvili\IEEEauthorrefmark{1}, Adam Doup\'e\IEEEauthorrefmark{1} and Gail-Joon Ahn\IEEEauthorrefmark{1}\IEEEauthorrefmark{4}} \\
\IEEEauthorblockA{ \IEEEauthorrefmark{1}Arizona State University,
\IEEEauthorrefmark{2}PayPal, Inc.
 \IEEEauthorrefmark{3}University of Tennessee, Knoxville, 
  \IEEEauthorrefmark{4}Samsung Research \\ 
\IEEEauthorrefmark{1}\{mbitaab, haehyun, aoest, penghui.zhang, zhibo.sun, rpourmoh, tbao, fishw, yans, doupe, gahn\}@asu.edu \\
\IEEEauthorrefmark{2}doowon@utk.edu
}}

\IEEEoverridecommandlockouts
\IEEEpubid{\makebox[\columnwidth]{978-0-7381-3261-7/20/\$31.00 ~\copyright2020 IEEE \hfill} \hspace{\columnsep}\makebox[\columnwidth]{ }}


\maketitle

\begin{abstract}\input{abstract}

\end{abstract}


\input{intro} 
\input{background}

\input{methodology}
\input{analysis}

\input{discuss}

\input{relatedwork} 
\input{conclusion}

\bibliography{references.bib}{}
\bibliographystyle{plain}


\end{document}

%% file: abstract.tex
As the COVID-19 pandemic started triggering widespread lockdowns across the globe, cybercriminals did not hesitate to take advantage of users' increased usage of the Internet and their reliance on it. In this paper, we carry out a comprehensive measurement study of online social engineering attacks in the early months of the pandemic. By collecting, synthesizing, and analyzing DNS records, TLS certificates, phishing URLs, phishing website source code, phishing emails, web traffic to phishing websites, news articles, and government announcements, we track trends of phishing activity between January and May 2020 and seek to understand the key implications of the underlying trends.

We find that phishing attack traffic in March and April 2020 skyrocketed up to 220\% of its pre-COVID-19 rate, far exceeding typical seasonal spikes. Attackers exploited victims' uncertainty and fear related to the pandemic through a variety of highly targeted scams, including emerging scam types against which current defenses are not sufficient as well as traditional phishing which outpaced the ecosystem's collective response.

%% file: intro.tex
\section{Introduction}
\label{s:intro}

The COVID-19 pandemic upended daily life across the globe and has led to unprecedented changes from two perspectives.
First, the ensuing widespread lockdowns, travel restrictions, and telecommuting arrangements (working from home) have significantly increased users' reliance on online services.
Second, continuous updates from news outlets and social media caused panic about the rapid spread and dangers of the disease~\cite{gorter2020impact}.
Unfortunately, this increased usage of the Internet and the unstable emotions of its users has left the users vulnerable to online social engineering attacks such as scams and phishing~\cite{abraham2010overview}.
For instance, attackers exploit users' fear to trick them into \textit{acting now} instead of making an informed decision:
For example, one COVID-19 phishing email exploits Internet users' fear by stating that \textit{``this is the last set of test kits.''} available for purchase.
Besides fear, attackers also capitalize on their victims' generosity:
Because many people desire to help others during major tragedies, scammers create fake donation campaigns as a lure to mount attacks.

Abundant news reports and government alerts about phishing attacks underscore the significance of anti-phishing systems~\cite{covidcybercrime}.
However, such reports are generally anecdotal, and comprehensive studies on phishing (and other cybercrime) related to the pandemic are needed to inform society to better respond to these threats.

This need, combined with a lack of studies on the relationship between large societal shifts (such as the pandemic) and phishing attacks, motivates us to investigate the effect of COVID-19 on phishing trends, the effects of these changing trends on phishing victims, and possible defenses that can be implemented or enhanced to protect users in this dangerous online landscape.
Specifically, in this paper, we seek to answer the following research questions:

\smallskip
\begin{itemize}
    \item
        How has the COVID-19 situation affected trends in phishing attacks?
    \item
        How many victims have visited phishing websites related to the pandemic?
    \item
        What are the attackers exploiting?
    \item
        How can we improve anti-phishing systems to protect users and organizations from phishing threats that leverage massive global situations like COVID-19?
\end{itemize}

\smallskip
To answer the research questions, we collected a variety of datasets in the course of conducting our research:
(1) we collected news articles and government announcements about phishing attacks related to the COVID-19 pandemic;
(2) we gathered and monitored DNS records, TLS certificate transparency logs, and phishing website reports to measure how the pandemic has affected the Internet infrastructure; 
(3) we crawled the source code of COVID-19-themed phishing websites from among the reported URLs to explore novel types of phishing content;
(4) by collaborating with a major financial services organization, we used a specialized network monitor to analyze trends in victim traffic to phishing websites and the volume of phishing reports by users of the organization. This gives us an unparalleled view from the organization's perspective; and
(5) we collected COVID-19-related discussions from two large underground forums to understand cybercriminals' objectives and activities related to the pandemic.


\smallskip
We performed a multi-faceted analysis of the collected datasets.
Through our analysis, we made several interesting findings about the early months of COVID-19:

\smallskip
\begin{itemize}
    \item{\textbf{Record-breaking attack volume.}}
    We observed that traffic to phishing websites reached record levels in March and April 2020, with up to 2.2 times more users falling victim to phishing than in the preceding months.
    Cybersecurity warnings from governments and major industry organizations lagged behind these attacks.
    
  \item{\textbf{Social engineering strategies.}}
    During COVID-19, attackers exploited both users' altruism and self-interest.
    For example, we found attacks that impersonated the Centers for Disease Control and Prevention (CDC) and harvested user credentials and identities while making users believe they were making a donation.
    Conversely, myriad fraudulent storefronts pretended to sell personal protective equipment (PPE) or attempted to sell counterfeit goods such as fake COVID-19 testing kits.

    \item{\textbf{Current defenses.}}
    Traditional anti-phishing systems are primarily reactive in nature and, thus, they struggle to quickly protect users, at scale, in the face of novel types of phishing attacks.  In addition, ecosystem defenses against non-phishing scams have a lesser degree of maturity.
\end{itemize}

\smallskip
Much to our surprise, despite historically being the most prevalent browser-based threat~\cite{safebrowsing}, phishing was not the most common threat among all COVID-19-related online attacks.
In the first four months of 2020, we identified 467,323 COVID-19-related domain registrations; a curated whitelist indicated that just 0.16\% (774) of these domains were benign~\cite{mispwarninglists}.
Among all the registered domains, we found out that only 0.22\% (1,047) of them appeared on phishing blacklists.
Therefore, we concluded that phishing websites only represented \emph{a small fraction} of malicious COVID-19 domains.
As such, defenses against other types of scams are as important as anti-phishing defenses.
To this end, we provide in our paper a taxonomy of other types of scams, such as fake storefronts or deceptive donation pages.
We also recommend new ecosystem defenses to identify these scam websites and protect users from them as future work.
The contributions of this paper are thus as follows:
\begin{itemize}
    \item Our study clearly shows that attackers move quickly to develop novel types of attacks to exploit users' increased vulnerability during a crisis.
    
    \item This work is the first step in comprehensively investigating phishing attack trends as a result of COVID-19 and motivates standardized approaches to not only keep up with the agility of attackers but also help guide timely mitigations to protect users from sophisticated online scam threats.
\end{itemize}



%% file: background.tex
\section{Background}
\label{s:backg}
In social engineering attacks, attackers lure victims to disclose sensitive
information. Phishing is a common type of social engineering attack in which criminals masquerade as trustworthy entities to take advantage of targeted victims. Attackers typically manipulate the victims to submit their
credentials by exploiting their fear, curiosity, charitable spirit, or
apprehension~\cite{vargas2016knowing, vishwanath2011people}. As more routine tasks become digital, people increasingly rely on the Internet. The
migration of tasks from traditional (e.g., paper-based) formats to online services has provided
opportunities for cybercriminals to lure victims~\cite{hiscox}. 

Cybercrime typically has three main components: (1) a victim who is the target of a cyber-attack, (2) a motive which is the criminal's incentive for committing the attack, and (3) a vulnerability or opportunity that enables the crime to take place~\cite{shinder2008scene}.
When the online presence of users increases, the first two conditions will be met.
Different principles affect the third factor for a successful attack, such as distraction, time pressure, compassion, and need~\cite{stajano2011understanding}.
When the aforementioned conditions are met, attackers usually ramp up their activities to maximize their success rate~\cite{22}.
In the past, attackers have seen natural disasters as a prime opportunity to carry out social engineering.
For example:

\begin{description}
\item[Ebola Virus Outbreak.]
The largest Ebola outbreak occurred in 2014 and lasted two years in west Africa. Although the Ebola virus did not spread worldwide, attackers targeted affected groups of people with phishing and scams. Barracuda Networks reported that 200,000 spam emails with Ebola news updates attempted to make people open malicious links, and 700,000 scam emails solicited donations to fictitious organizations~\cite{korolov_2014}.

\item[Australia's Bushfire.]
During the Australian bushfire that happened in late 2019, attackers claimed to be from large organizations, the government, or popular charities to deceive people into donating money or providing sensitive information~\cite{bushfire}.
\end{description}

Unlike the disasters above, the COVID-19 pandemic has caused worldwide panic and, thus, miscreants have been exploiting the empathy and fear of people on a larger scale, in part by deploying scams and phishing websites with COVID-19-themed content~\cite{Bewarescams}. 


%% file: methodology.tex
\section{Dataset}
\label{s:method}
In this section, we discuss the datasets that we collected to enable us to investigate social engineering scams related to the pandemic in the remainder of the paper.
Our datasets cover news reports and government announcements, domain names, TLS certificates, reported phishing URLs, reported phishing emails, and posts from underground forums.
\autoref{tab:dataset_stats} shows an overview of these datasets.

\subsection{Terminology}

For the sake of brevity, throughout the paper, we use the term \emph{corona-related} to refer to any data related to the COVID-19 pandemic.

To find corona-related news and domains in our datasets, we used keywords that include: \textit{covid}, \textit{covid-19}, and \textit{coronavirus}, along with their permutations.
Using regular expressions, we also considered keywords related to COVID-19 that use special characters or numbers, such as \textit{c0-vi-d-19}. 

\begin{table*}[t]
    \centering
    \begin{tabular}{rccr}
        \toprule
        \textbf{Data Content} & \textbf{Data Source} & \textbf{Date Range} & \textbf{Number of Samples} \\
        \midrule
        News, Government, and Companies announcements & Google News & 01/01/20 -- 05/12/20 & 756\Bstrut\\
        Domain names & RiskIQ, Domaintools, WhoisDS & 01/01/20 -- 04/30/20 & 467,323\Bstrut\\
        TLS certificates & Google Rocketeer CT log & 01/01/20 -- 04/30/20  & 33,596,126\Bstrut\\
        Reported phishing URLs & APWG, OpenPhish & 06/01/19 -- 04/30/20 & 3,191,012\Bstrut\\
        Source code of phishing websites & APWG & 01/01/20 -- 04/30/20 & 49,306\Bstrut\\
        Reported phishing emails & Financial services provider & 01/01/20 -- 04/30/20 & 387,251\Bstrut\\ 
        Victim web traffic to phishing websites & Financial services provider & 01/01/20 -- 04/30/20 & Not disclosed\Bstrut\\ 

        Posts from underground forums & Nulled.to, Cracked.to & 01/25/20 -- 05/06/20 & 3,530\Bstrut\\
        \bottomrule
    \end{tabular}
    \caption{An overview of our datasets.}
    \label{tab:dataset_stats}
\end{table*}

\subsection{Summary of The Dataset} 

We collected the following data for this study:

\begin{description}
  \item[News and government announcements.]
      From media outlets, governments, and private companies, we automatically collected news and announcements about social engineering attacks related to the pandemic.
 We then filtered those that are relevant to phishing attacks that reference the pandemic. We searched Google News with the keywords $\{$\textit{corona, covid-19, scam, phishing}$\}$ to gather news from both government and non-government websites.

    \item[Domain names.]
To investigate changes in domain registration trends, we collected DNS records from three different sources: Domaintools~\cite{domaintools}, Whois Domain Search~\cite{whoisds}, and RiskIQ~\cite{covidcybercrime} to find corona-related domain names registered daily. Both Domaintools and RiskIQ provide filtered lists of registered corona-related domains, whereas Whois Domain Search provides all daily domain registrations which we then retrieved and scanned for corona-related domains.

    \item[TLS certificates.]
To find certificates issued to web sites with corona-related domain names and phishing websites related to the pandemic, we collected 144,590,199 TLS certificates using the Google Rocketeer CT  log~\cite{googldrocketeer}.

    \item[Reported phishing URLs.]
We collected phishing URLs submitted to OpenPhish~\cite{openphish} and the Anti-Phishing Working Group (APWG)~\cite{apwgsite}.

    \item[Source code of phishing websites.]
We crawled the source code (i.e., page content) of 49,306 phishing websites between January 2020 and April 2020 (using the APWG URLs) to investigate corona-related phishing content and techniques.

    \item[Phishing emails and traffic to phishing websites.]
    Between January and April 2020, we analyzed 387,251 phishing emails reported by users, and signals based on victim traffic to phishing websites, by collaborating with an organization commonly targeted by phishing.

    \item[Underground forums.]
To understand shifts in criminals' activities amid the COVID-19 pandemic, we crawled corona-related discussions in two popular underground forums: \emph{Nulled.to} and \emph{Cracked.to}, which have more than 2.8 million and 1.1 million registered members, respectively.

\end{description}

%% file: analysis.tex
\section{Measurement Results}
\label{s:anal}
In this section, we first discuss overall measurement results based on the combination of our datasets, and we then present detailed findings from each individual dataset.
In \autoref{ss:domcerts}, we show trends in corona-related DNS records and reported phishing websites. Next, in \autoref{ss:comms}, we study news from both private and government news outlets to understand the perceived importance of COVID-19 themed phishing and scams. We follow with an analysis of real victim traffic to phishing websites in \autoref{ss:phishingtrends}. Then, we categorize and explain various types of COVID-19-themed phishing websites in \autoref{ss:covidphishing}. In \autoref{ss:undermarket}, we use the underground forum data to characterize the corona-related topics discussed by criminals.


\subsection{General Findings}
\label{ss:summary1}
\paragraph{Record-breaking victims}
Even though in our dataset, phishing attacks leveraging the pandemic seem to be negligible compared to traditional attacks, 
there was a record-breaking number of overall phishing victims during this period (described fully in \autoref{ss:phishingtrends}).
We believe this is because phishing lures effectively exploited their victims' pandemic-related concerns, and that attackers caught victims off guard with high-quality phishing websites (\autoref{ss:covidphishing}).
We observed that the number of news reports and government announcements regarding corona-related phishing attacks increased rapidly from March 2020 (\autoref{ss:comms}).
However, there were still many victims after that, which implies that typical anti-phishing systems' reactive mitigation strategy is not enough to protect users from a surge of novel phishing attacks.

\paragraph{No ecosystem defenses for non-phishing scams}
We found 467,323 new corona-related domain names and 17,699 new certificates issued to corona-related domains from our observation period. 
Only 0.22\% (1,047) of these new corona-related domains were reported to phishing blacklists (\autoref{ss:domcerts} and \autoref{ss:phishingtrends}).
Also, one curated list suggests that only 774 of the domains hosted legitimate websites~\cite{mispwarninglists}.
Furthermore, a recent FTC report showed that 54,813 corona-related scams reported across the U.S. from January led to \$40.13M in loss due to fraud~\cite{covidscam_jun11}. Based on this report, online shopping is the most commonly reported scam.
This also implies that phishing is just one type of many corona-related attacks.
Other corona-related domains can be used for different fraudulent purposes, such as non-phishing scam websites or e-mail spam.
Cybercriminals exploit the ecosystem's lack of defenses against such non-phishing scams. 


    
    


\begin{figure*}[t]
    \centering
    \includegraphics[width=0.99\linewidth]{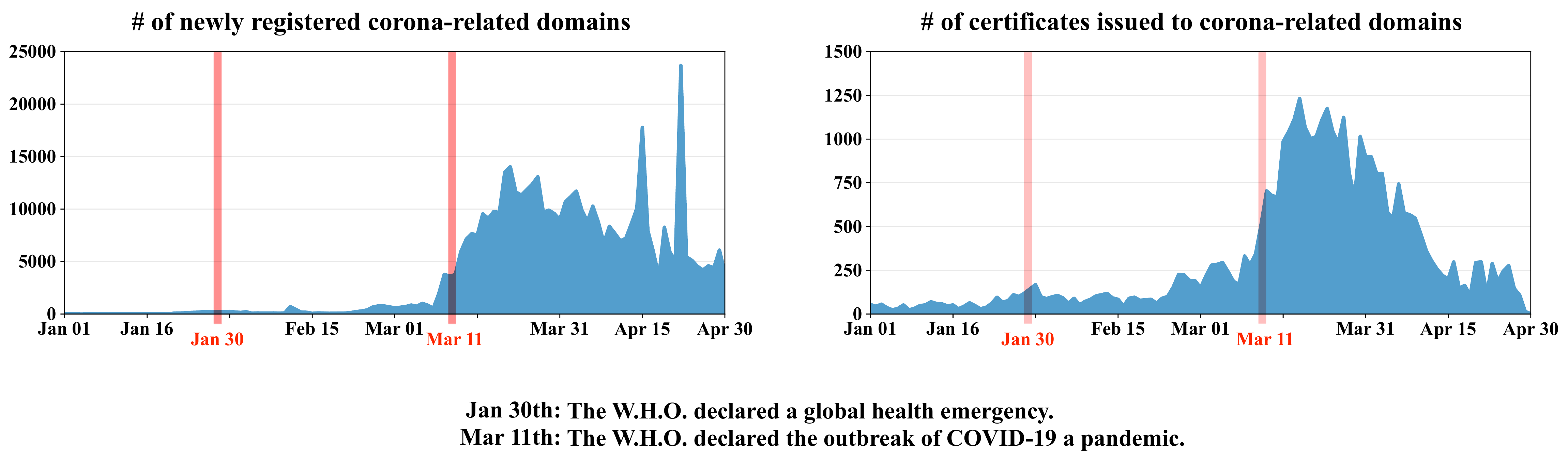}
    \caption{The number of newly issued certificates used for corona-related domains and the number of newly registered corona-related domains.}
    \label{f:coronadomains}
\end{figure*}


\subsection{Domain Names and Certificates}
\label{ss:domcerts}
We provide an overview of corona-related websites and phishing sites based on domain names and URLs.

As \autoref{f:coronadomains} illustrates, we found that the number of corona-related domain names started increasing significantly from early March 2020.
On average, 155 corona-related domain names were registered every day before March 2020; however, this number increased to 7,453 in March and April. 
With the increase in new corona-related websites, it is difficult to distinguish between legitimate and malicious websites.
As shown in~\autoref{f:coronadomains}, the number of certificates issued to corona-related domain names increases starting in February 2020 and peaks in March 2020.

We further analyzed 72 certificates of corona-related HTTPS phishing websites as in \autoref{t:apwgnumbers}.
\autoref{t:phishingcerts} shows that CAs which issued certificates to corona-related phishing websites with the number of certificates and revoked certificates.
Except for GoDaddy and Sectigo, the other CAs use the Automatic Certificate Management Environment protocol, allowing attackers to obtain TLS certificates easily.
Through May, only GoDaddy and cPanel revoked certificates issued to phishing websites (five and two certificates, respectively). 

We queried the collected DNS records against the APWG, OpenPhish, GSB, and RiskIQ blacklists to estimate how long it takes for a corona-related phishing URL to be detected after launch.
\autoref{tab:inters} shows the number of intersections between collected corona-related DNS records and each phishing blacklist.
We calculate the \emph{average gap} between the registration date and date of each report.
From $467,323$ DNS records we collected, $110$ domains were reported to APWG, on average $3.6$ days after registration, and $72$ domains were reported to OpenPhish $4.8$ days after registration.
As GSB and RiskIQ do not provide the date that blacklisted domains were reported, we were unable to calculate the average gap.
Also, we found that 110 domains reported to the APWG were newly registered in 2020; however the other 44\% of reported domains existed before January $1^{st}$ 2020.
\autoref{t:apwgnumbers} shows that 198 domains were reported to the APWG from January 2020 to April 2020.


\begin{table}[t]
    \centering
    \begin{tabular}{rrr}
        \toprule
        \textbf{Month} & \textbf{\# of Reported URLs} & \textbf{\# of HTTPS Domains}\\ 
        \midrule
        January 2020 &   0 &  0\\
        February 2020 &   5 &  1\\
        March 2020 & 171 & 37\\
        April 2020 & 140 & 34\\
        \multicolumn{1}{r}{Total} & 316 & 72\\
        \bottomrule
        \multicolumn{3}{r}{\# of unique domains among the reported URLs = 198.} \\
    \end{tabular}
    \caption{The number of corona-related URLs reported to the APWG per month and the number of HTTPS domains among the reported URLs.}
    \label{t:apwgnumbers}
\end{table}

\begin{table}[t]
\centering
    \begin{tabular}{lcrr}
    \toprule
    \multicolumn{1}{c}{\textbf{CAs}} & \textbf{ACME} & \textbf{\# of Certs.} & \textbf{\# of Revoked Certs.} \\
    \midrule
    Let's Encrypt & \tick & 31 & 0\\
    cPanel & \tick & 22 & 2\\
    Go Daddy & \cross & 8 & 5\\
    Cloudflare & \tick & 6 & 0\\
    Sectigo & \cross & 6 & 0\\
    \bottomrule
    \end{tabular}
\caption{CAs, the number of certificates, and the number of revoked certificates that were used for phishing.}
\label{t:phishingcerts}
\end{table}

\begin{table}[t]
    \centering
    \begin{tabular}{rrr}
    \toprule
    \textbf{Blacklists} & \textbf{\# Intersections} & \textbf{Avg. Gap (days)}\\ 
    \midrule
    APWG      & $110$ & $3.6$\\
    OpenPhish &  $72$ & $4.8$\\
    RiskIQ-BL & $316$ & N/A\\
    GSB       & $833$ & N/A\\ 
    \bottomrule
    \multicolumn{3}{r}{\# of unique domains among the blacklisted domains = 1,047.}
    \end{tabular}
    \caption{Number of intersections and average gap for anti-phishing entities' blacklists}
    \label{tab:inters}
\end{table}


\subsection{Public Phishing Guidance}
\label{ss:comms}
To study the effect of the COVID-19 pandemic on corona-related social engineering attacks, we examined how news outlets, governments, and large companies have provided guidance or warnings against phishing and scam attacks.
To this end, we collected daily news reports and government announcements starting from January 2020.

\begin{figure}[t]
    \centering
    \includegraphics[width=0.8\linewidth]{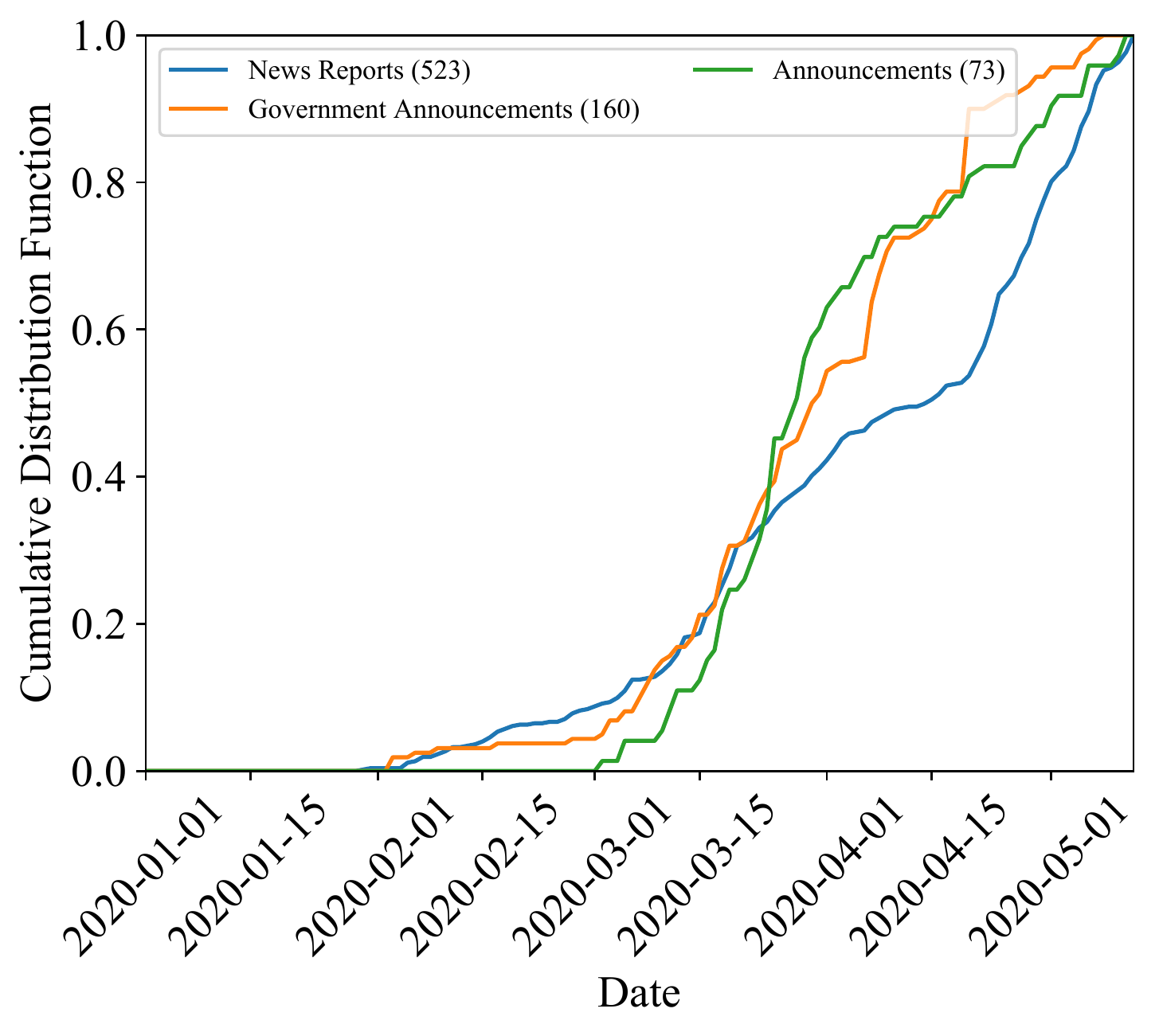}
    \caption{The number of news and government announcements about corona-related phishing attacks.}
    \label{fig:GovernmenandNews}
\end{figure}

\autoref{fig:GovernmenandNews} shows the number of corona-related news reports and government and company announcements.
The news reports began to reference corona-related scams on January $30^{th}$, stating that several phishing campaigns were sending corona-related emails containing malware.
The first official U.S. government announcement about corona-related scams was made on February $4^{th}$ by the U.S. Securities and Exchange Commission (SEC), which warned people about cybercriminals trying to leverage the COVID-19 situation.

After the original SEC announcement, the number of news reports related to corona-related scams increased rapidly. However, the U.S. government did not make many announcements about corona-related phishing attacks until the beginning of March 2020, when several additional government departments posted alerts.  As shown in \autoref{fig:GovernmenandNews}, companies started to directly address corona-related scams at the beginning of March, with an increasing trend thereafter. The rapid growth of news from different sources motivates us to further investigate these scams.

\begin{figure}[t]
    \centering
    \includegraphics[width=0.8\linewidth]{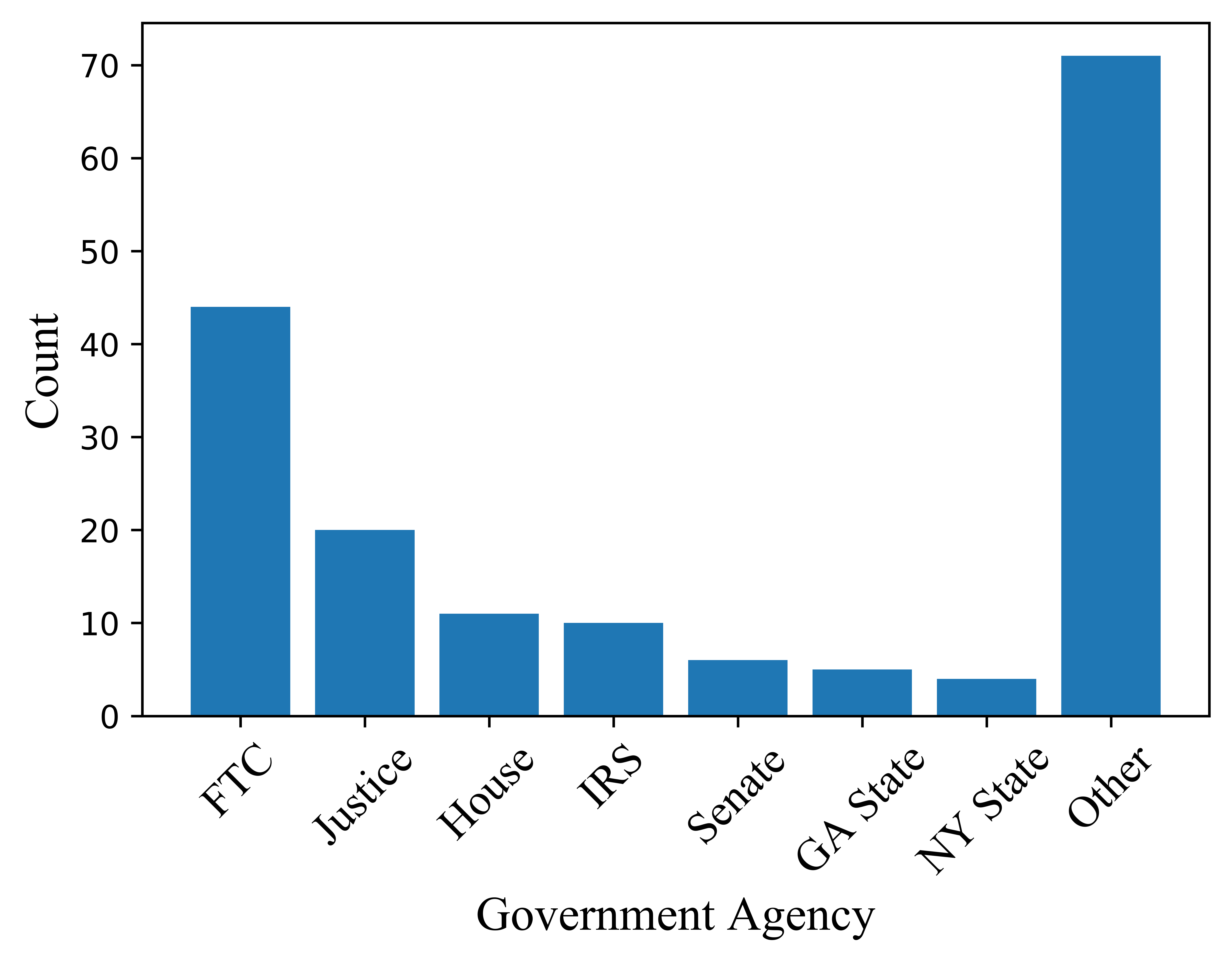}
    \caption{Government announcements regarding corona-related phishing and scams, grouped by government agency.}
    \label{fig:gov_news}
\end{figure}

As illustrated in \autoref{fig:gov_news}, most of the U.S. government announcements are from FTC, followed by the Department of Justice, while states only published one warning on their official websites (shown as ``other''). 
The announcements warn people about scams and the threat of the theft of sensitive information such as the Social Security Number (SSN), and typically include detailed guidance for how members of the public can protect themselves from fraud.

\subsection{Phishing Trends}
\label{ss:phishingtrends}

\begin{figure}[t]
    \centering
    \includegraphics[width=0.78\linewidth]{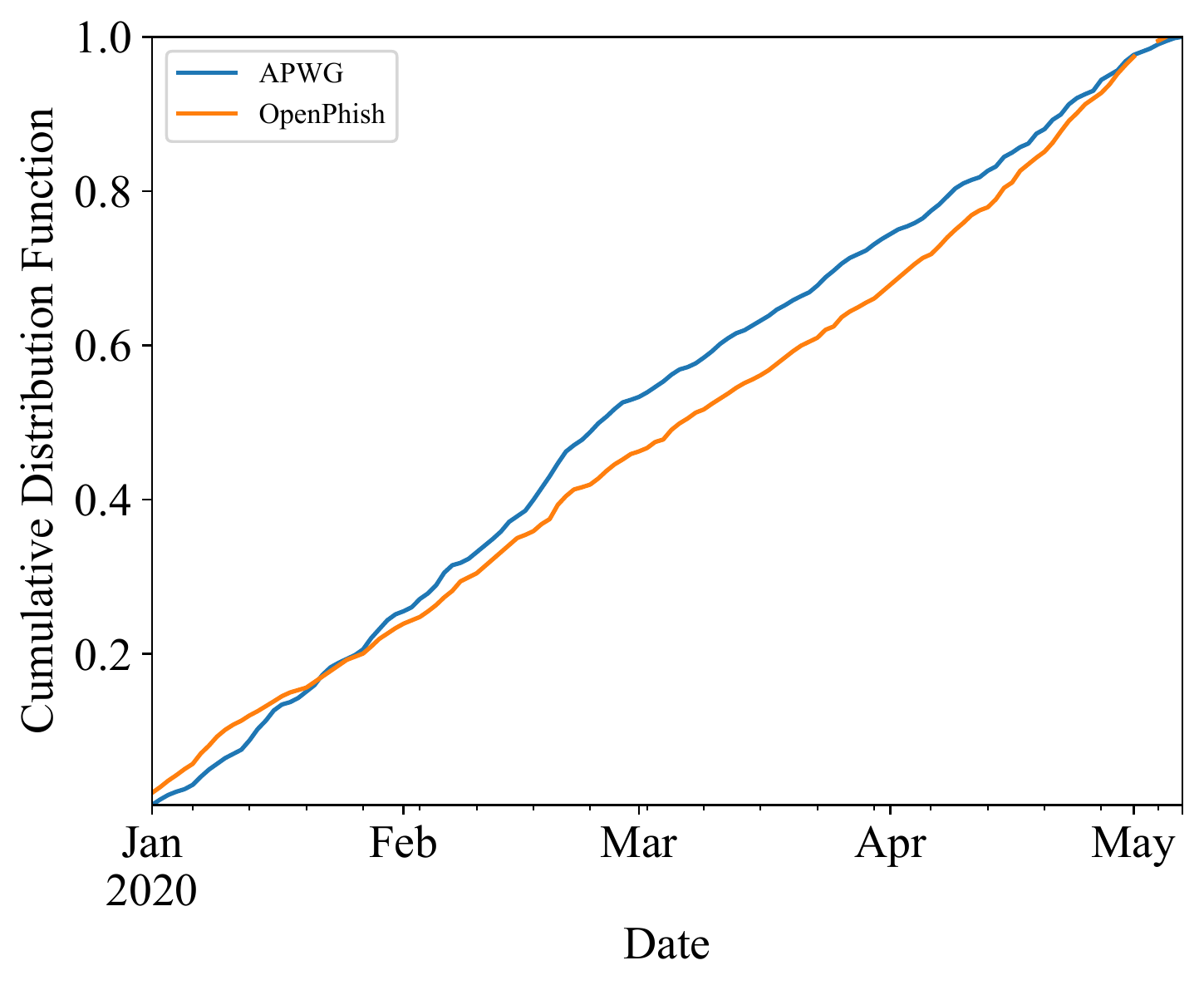}%
    \caption{CDF of unique phishing hostnames reported to two major anti-phishing entities.}
    \label{fig:stats}
    
\end{figure}

\begin{figure}[t]
\begin{subfigure}{.5\textwidth}
  \centering
  \includegraphics[width=.7\linewidth]{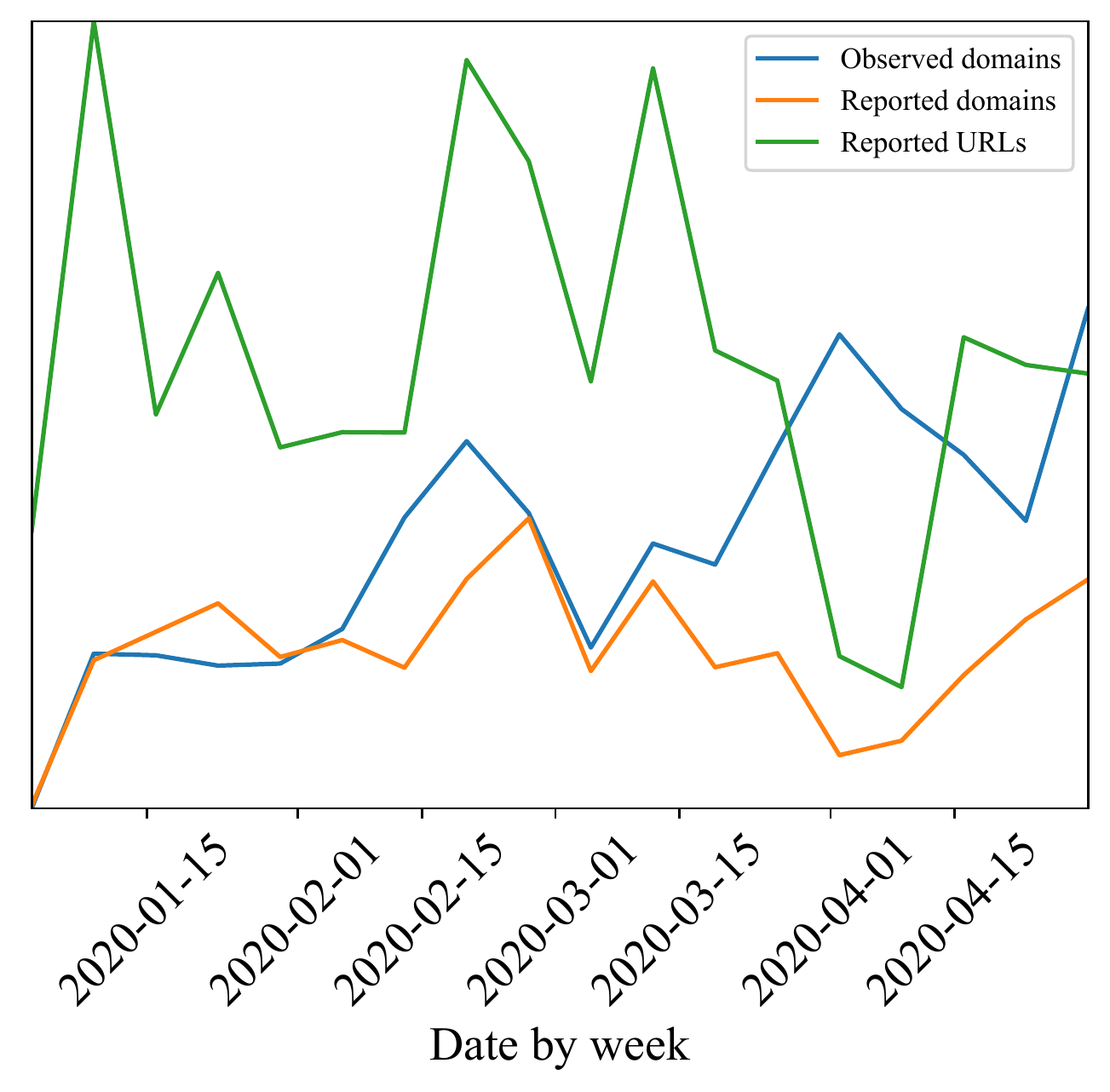}  
  \caption{Number of observed/reported phishing domains and URLs.}
  \label{fig:sub-first}
\end{subfigure}
\begin{subfigure}{.5\textwidth}
  \centering
  \includegraphics[width=.7\linewidth]{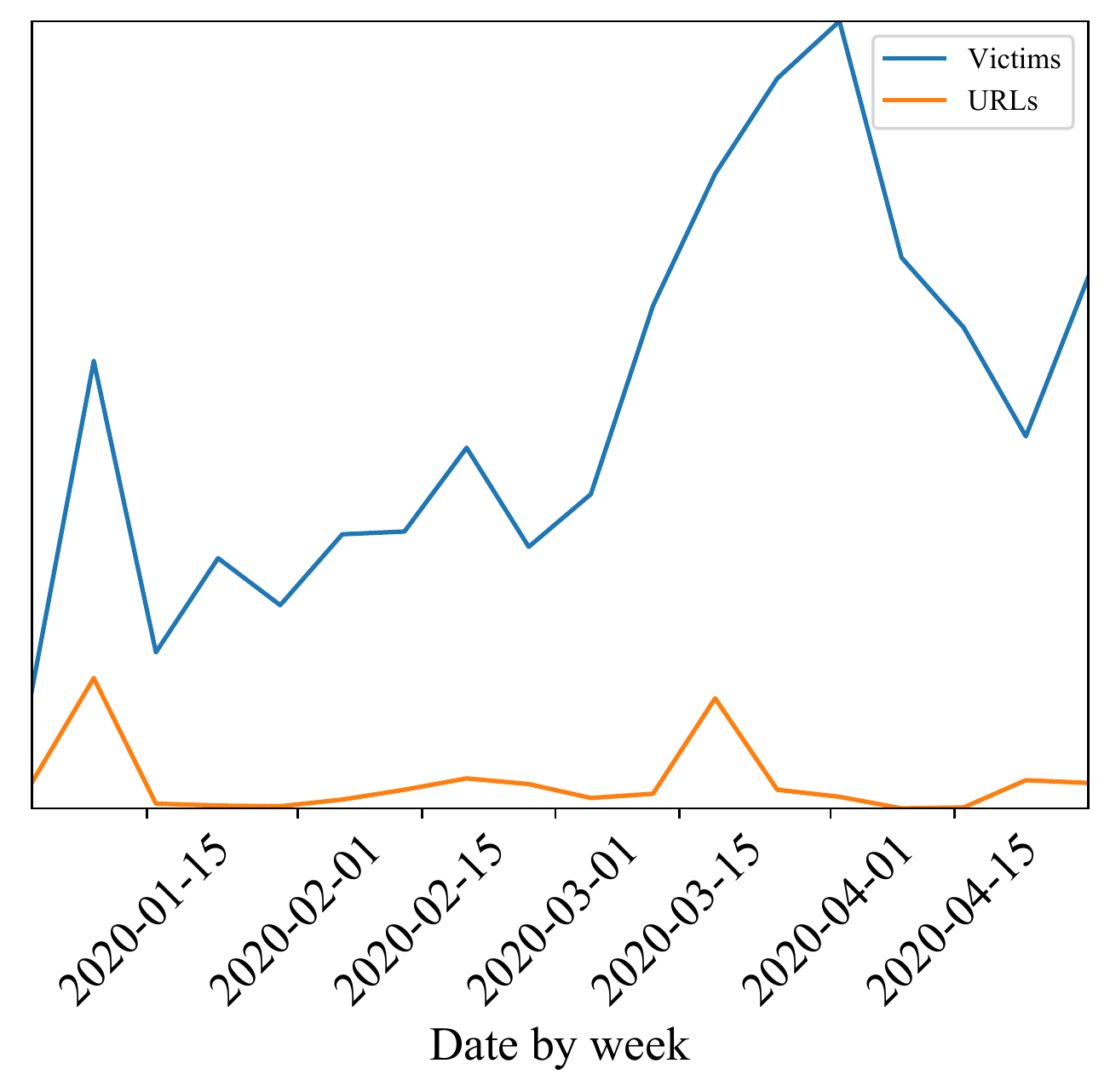}  
  \caption{Victim visits to phishing websites.}
  \label{fig:sub-second}
\end{subfigure}
\caption{Number of potential phishing victims identified by our network monitor, compared to the number of corresponding unique phishing URLs.  Victim counts increased substantially in March and April despite a lack of a significant increase in URLs.}
\label{fig:networkmonitor}
\end{figure}

As shown in \autoref{fig:stats}, the number of phishing hostnames reported to two major clearinghouses of phishing URLs did \emph{not} increase significantly during the COVID-19 outbreak.
However, hostname counts alone fail to accurately reflect the damage caused by phishing attacks, as certain high-impact websites may receive substantially more traffic than others as a result of their increased spamming activity or the ability to evade defenses~\cite{oest2020sunrise}. 

To deepen our insight into phishing trends during the crisis, we collaborated with a financial service organization that is commonly targeted by phishers and analyzed two additional datasets: (1) traffic to phishing websites by targeted victims 
and (2) phishing emails reported by users. The phishing website traffic was collected using a recently proposed network monitoring approach~\cite{oest2020sunrise} which passively measures victim visits to live phishing websites based on signals (referrer headers as well as third-party resources embedded on phishing websites) detectable by the organization. Specifically, it first analyzes web traffic logs to find events of interest. Then, after filtering out benign events, it looks for events correlated with known phishing URLs. Further analysis of the events enables us to determine how many victims have fallen for corona-related phishing websites\footnote{
    The network events recorded by this approach have a high probability of being linked to victims successfully fooled by phishers, and have been de-duplicated to reflect individual sessions.}.
As shown in \autoref{fig:networkmonitor}, the network monitor recorded a surge in phishing victims from late March; attack volume remained elevated throughout April. Overall, the total number of observed phishing victims in March and April was 2.207 and 1.706 times higher than in February, respectively, and also 2.165 and 1.674 times higher than in January, when the organization typically sees elevated phishing volume due to holiday shopping. 



Phishing reports that users sent to the organization validate our observed increase in victim traffic, as a sustained rise in reporting is directly linked to an increase in spamming~\cite{google_report_spam}. 1.06 times more emails were reported to the organization in March, and 2.64 times more in April, compared to the number of reports in February.

Interestingly, within this dataset, only 0.51\% of phishing websites had corona-related content.  Similarly, 0.02\% of emails had COVID-19 keywords in the title or body, while 0.43\% had such keywords in the sending email address.

The World Health Organization (WHO) declared a global pandemic on March $11^{th}$. Even though there were not many corona-related phishing websites, the number of victims increased dramatically after the WHO's pandemic announcement, as shown in~\autoref{fig:sub-second}. Moreover, the government announcements increased \textit{after} the victim counts reached their peak in March. We suspect that not many corona-related phishing websites could take advantage of many victims within a short period after March $11^{th}$. 


We conclude that through an increase in spamming activity against a larger attack surface, the pandemic led to record numbers of phishing victims. However, in the case of this organization, phishing attacks that leveraged COVID-19 as a lure were negligible compared to traditional attacks that simply impersonate the brand. We note that the latter trend may be skewed by the brand's industry sector, however. 

\begin{figure}[t]
    \centering
    \begin{subfigure}{0.51\linewidth}
        \centering
            \includegraphics[width=\textwidth]{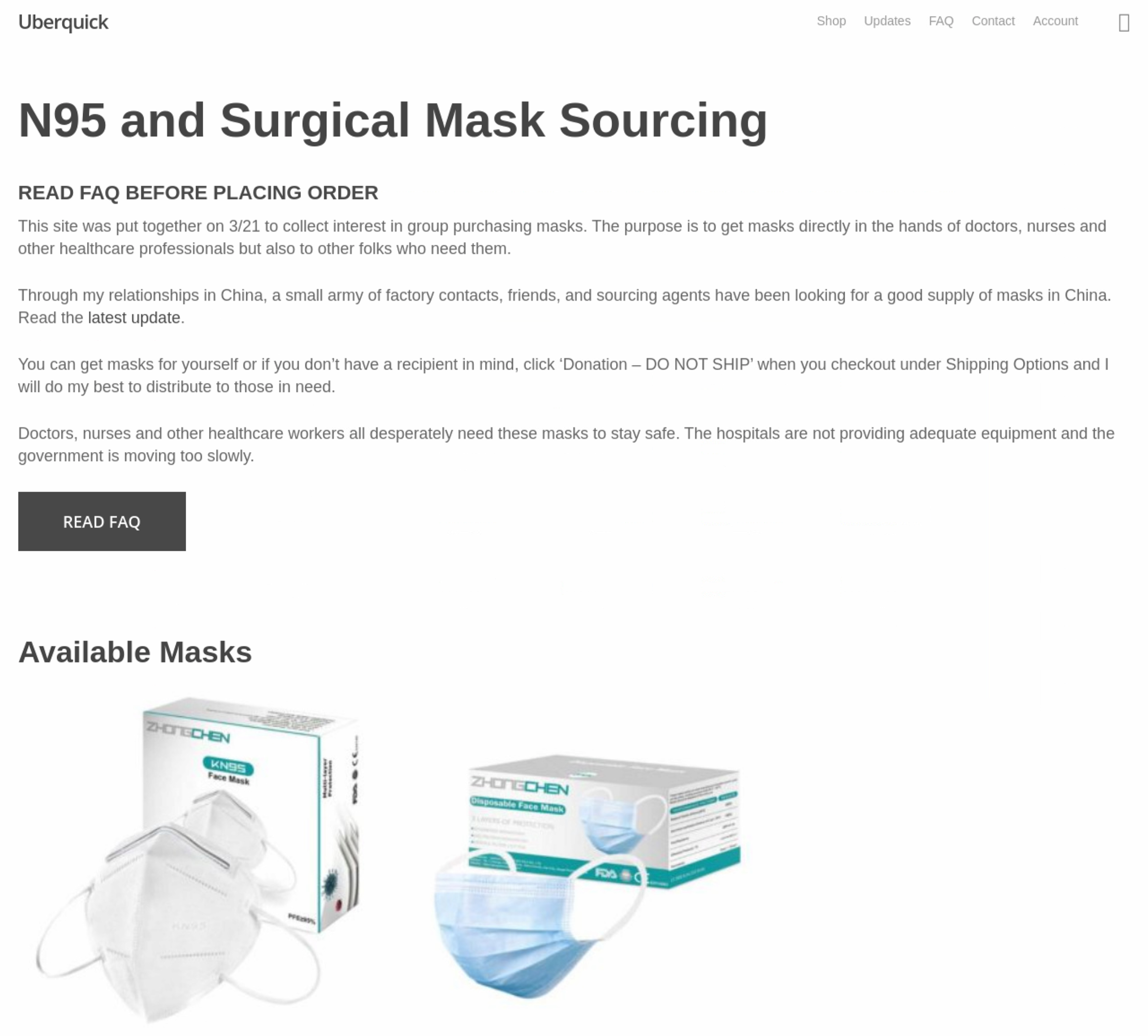}%
        \caption{Fraudulent PPE store.}
        \label{fig:ppe_phishing}
    \end{subfigure} 
    \hspace*{\fill}
    \begin{subfigure}{0.42\linewidth}
        \centering
            \includegraphics[width=\textwidth]{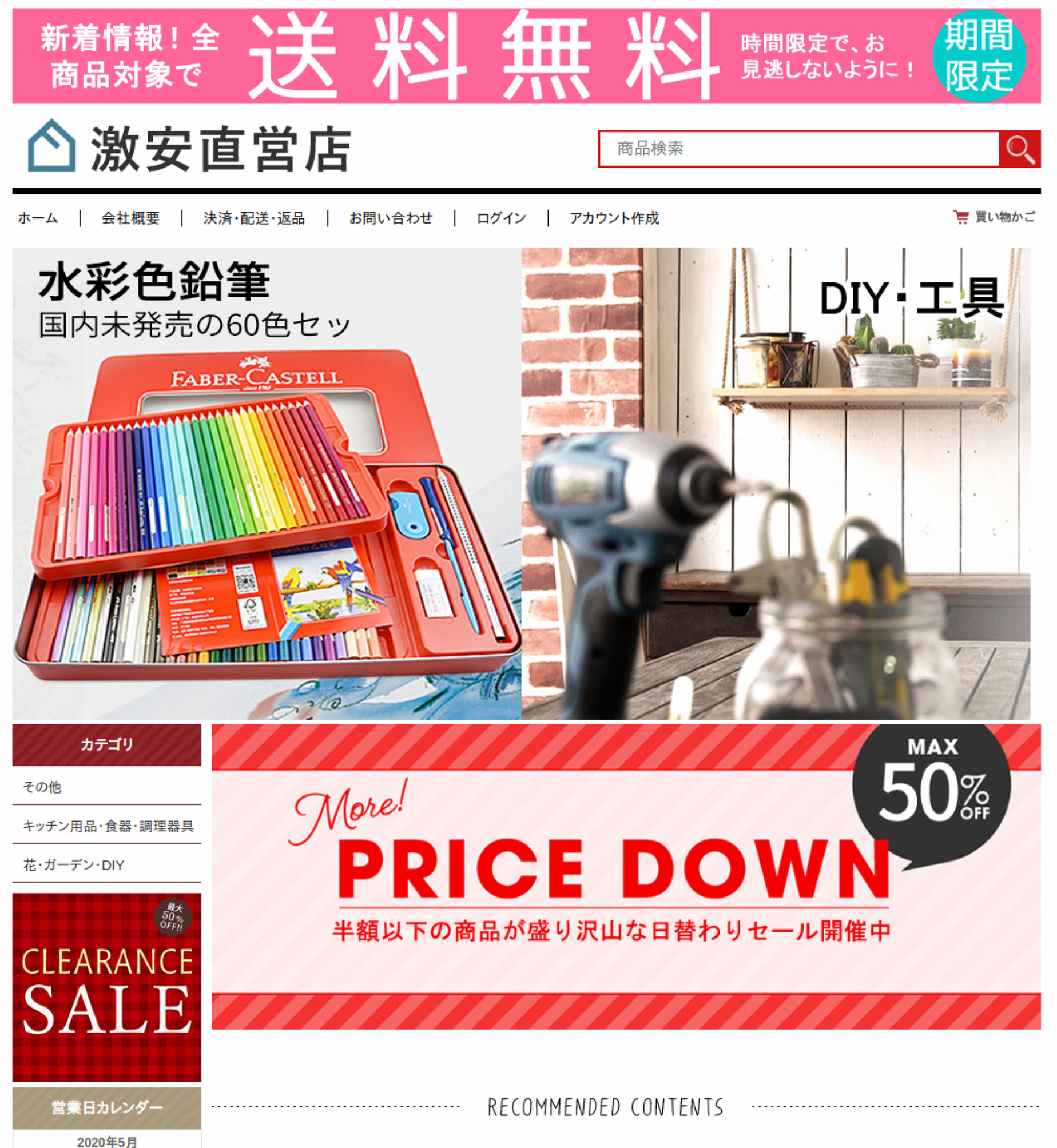}
        \caption{Free shipping fraudulent.}
        \label{fig:freeshipping_phishing}
    \end{subfigure} 
    
    \caption{Fraudulent website selling PPE and free shipping fraudulent store.}
    \label{fig:ppe_freeshipping_phishing}
\end{figure}

\begin{figure}[t]
    \centering
    \begin{subfigure}{0.43\linewidth}
        \centering
            \includegraphics[width=\textwidth]{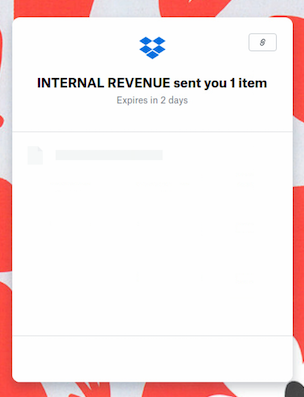}%
        \caption{Dropbox themed.}
        \label{fig:drobox}
    \end{subfigure}
    \hspace*{\fill}
    \begin{subfigure}{0.43\linewidth}
        \centering
            \includegraphics[width=\textwidth]{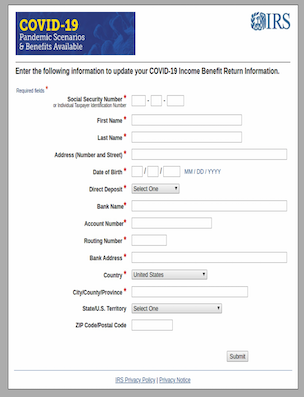}
        \caption{IRS themed.}
        \label{fig:irs}
    \end{subfigure}
    
    \caption{Phishing websites exploiting IRS.}
    \label{fig:irs_phishing}
\end{figure}

\subsection{COVID-19 Themed Phishing}
\label{ss:covidphishing}
We crawled and retrieved the source code of 49,306 phishing URLs from the APWG feed. We further considered corona-related scam websites reported to \textit{scam.directory}.
By analyzing their source code and page content, we found 255 unique phishing and scam websites with corona-related content.  

\smallskip
\noindent
\textbf{Donation-Themed Phishing.}
Some phishing websites steal sensitive information while making users think they are making a donation to a charity. For example, we identified one prominent phishing website which copied the look-and-feel of a donation portal run by a major organization. The phishing website deceptively informed visitors that they were making a donation to the ``CDC Response to CoronaVirus" through a registered charity, and provided detailed information about this (legitimate but unrelated) charity.  Victims successfully fooled by the attack---which we observed across several distinct domains---would think they were making a small donation to support the charity.  However, the phishing website would instead steal the victims' account credentials and credit card numbers, rather than processing actual donation payments.

\smallskip
\noindent\textbf{PPE Sale.}
Personal protective equipment (PPE), such as face masks and gloves, is in high demand when the public and healthcare workers try to protect themselves from communicable diseases.
In the early months of the pandemic, such equipment was also in short supply (and/or excessively priced) at major online retailers such as Amazon, eBay, or Walmart.
Therefore, people in urgent need of PPE may turn to other, less trustworthy sources.
To exploit the high demand, attackers designed fake shopping websites that sell PPE; an example is shown in \autoref{fig:ppe_phishing}. 

Several such websites embedded up-to-date corona-related information (e.g., COVID-19 statistics) in an effort to appear more legitimate. They lure people who require such equipment and then either steal their credentials, steal their money (without shipping any items), or sell them low-quality PPE. 



\smallskip
\noindent
\textbf{Fraudulent Online Shopping Websites.}
Fraudulent websites try to keep up with the look and feel of legitimate websites. As more and more legitimate organizations started to inform their customers about pandemic-related matters such as policy updates, new features, and COVID-19 statistics, attackers also included such information on their websites.
Similarly, some attackers advertised misleading ``free shipping'' offers on their fake shopping websites. Free shipping offers increase online sales and help attract visitors~\cite{freeshipping}. \autoref{fig:freeshipping_phishing} shows an example of such website.



\smallskip
\noindent\textbf{Exploiting Corona-related Events.}
Phishers not only generate corona-related phishing websites, but they also exploit other events related to the pandemic.
For example, to help address widespread financial hardship, the U.S. government offered stimulus funds by either direct deposit or a paper check.
However, different groups of people received payments at different times.
When people who received a check and shared this on social media, others might start to worry about if and when they also could receive their funds.  Phishers were quick to disguise themselves as the IRS to steal the personal information of people looking for the status of their stimulus payments.
For example, in \autoref{fig:irs_phishing}, IRS phishing websites steal Dropbox credentials or SSNs from users amid the pandemic. The text ``Expires in 2 days'' in the phishing website from \autoref{fig:drobox} conveys urgency so that visitors are more willing to open it.
\autoref{fig:irs} acquires users' SSN by declaring that they need such Personally Identifiable Information (PII) to process stimulus payments.

    

\subsection{COVID-19 in Underground Forums}
\label{ss:undermarket}
As underground forums are a key rendezvous point for cybercriminals~\cite{sun2018understanding,sun2021having}, we studied data from such forums to answer two questions: 
(1) Are corona-related topics popular in underground forums? 
(2) What do members discuss about corona-related topics? 

First, we measured the number of threads and posts within a one-week sliding window, as these are indicative of new discussion topics and members' overall levels of activity, respectively.
We found 2,913 members engaging in 144 coronavirus-related threads among 3,530 posts. 
Then, we manually analyzed all of these threads and 
categorized them into four topics:
(1) account (54.7\%, e.g., selling compromised accounts);
(2) virus (34\%, e.g., what the coronavirus is); 
(3) money (7.8\%, e.g., how to make money during the pandemic); 
(4) hacking (3.5\%, e.g., setting up COVID-19 related phishing sites).
Members in underground forums predominantly discussed compromised accounts and the coronavirus itself, as shown in Table~\ref{tab:content_category}.

\begin{figure}[t]
    \centering
    \includegraphics[width=.95\linewidth]{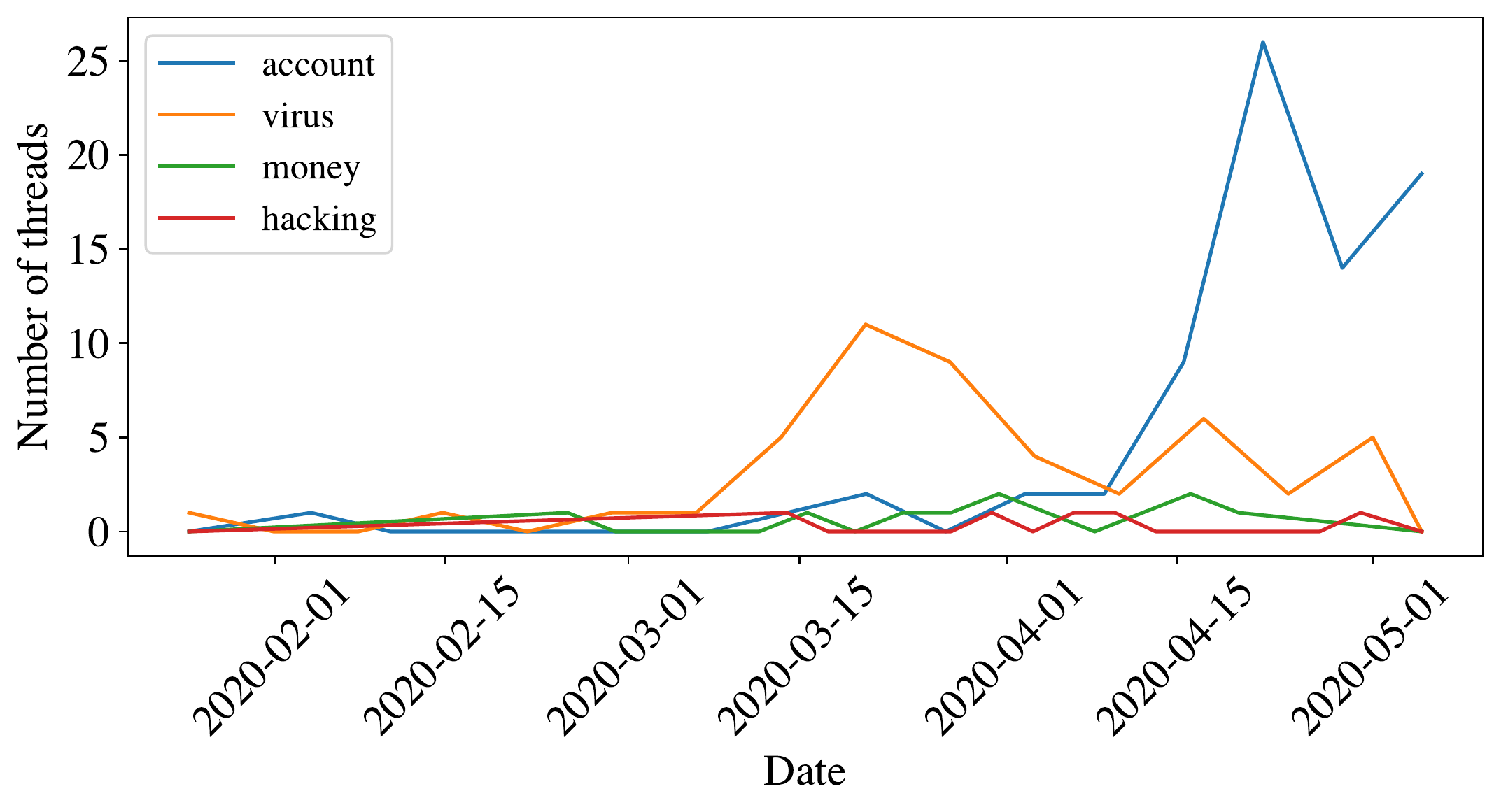}
    \caption{Topics discussed in underground forums.}
    \label{fig:content_time}
\end{figure}

\autoref{fig:content_time} shows the number of different topics discussed in underground forums over time.
At the beginning of the pandemic, people mainly discussed and shared information on the coronavirus; discussions shifted to compromised accounts from April $1^{st}$, and there is a significant peak at April $15^{th}$.
58.3\% of the new threads about selling compromised accounts were started during this period.
Because the total number of observed phishing victims peaked between March $1^{st}$ and April $15^{th}$ (as shown in \autoref{fig:networkmonitor}), we suspect that the attackers posted about stolen accounts after they successfully launched their attacks.

\begin{table}[t]
\centering
\footnotesize
    \begin{tabular}{crr}
    \toprule
    \multicolumn{1}{c}{\textbf{Content Category}} & \textbf{Sub-category} & \textbf{\% of discussions}\\
    \midrule
    
    \multirow{2}{*}{Virus} & General Discussion & 31.9\%\\
     & Virus Protection &  2.1\%\\
     \cdashline{1-3}[0.8pt/2pt]
     
     \multirow{5}{*}{Account} & Game Accounts &  20.6\%\\
     & VPN Accounts &  13.5\%\\
     & Video Site Accounts &  9.9\%\\
     & Music Site Accounts &  5.7\%\\
     & Porn Site Accounts &  5.0\%\\
     \cdashline{1-3}[0.8pt/2pt]
     
  \multirow{2}{*}{Money} & General Discussion  & 3.5\%\\
    & Making Money via E-whoring &  4.3\%\\
    \cdashline{1-3}[0.8pt/2pt]
    
    \multirow{3}{*}{Hacking} & Cracking &  1.4\%\\
    & Account Checkers &  1.4\%\\
    & Phishing &  0.7\%\\
    
    \bottomrule
    \end{tabular}
\caption{Coronavirus-related discussion in underground forums. }
\label{tab:content_category}
\end{table}

    
     
     
    
    

%% file: discuss.tex
\section{Discussion}
\label{s:discuss}

Our dataset and analysis represent a current snapshot of specific corona-related cybercrime and shed light on how attackers exploited the COVID-19 pandemic in the early months of the outbreak. The pandemic upended daily lives across the globe and, consequently, resulted in unprecedentedly rapid changes across the digital world. These changes open up a new yet important set of challenges for website owners, users, governments, and security researchers to be able to adapt accordingly. 

Our results demonstrate that attackers remain several steps ahead of typical modern anti-phishing defenses and will take advantage of a global crisis to directly harm online users.
It is of great importance to collaboratively deploy anti-phishing systems that can better adapt to changes in the ecosystem, to narrow the attack window available to phishers and perhaps go further to offer proactive defenses. 

From the corona-related domain perspective, 0.16\% of the domains in our dataset were known to be benign, and 0.22\% of the domains were known to be malicious.
We discovered that many of the other unknown domains are used for non-phishing scams such as fake storefronts, which can harm users, yet remain out of the reach of traditional anti-phishing defenses.

From our deep dive into the content of corona-related phishing websites, we noted the importance of human factors and how attackers exploit individuals' pandemic-driven wants and needs.
Hence, it is critical to raise awareness so that users better understand social engineering attacks and have access to the appropriate resources to protect themselves when technical mitigations fail to do so.

This paper is the first step toward investigating phishing attacks, trends, and consequences amid a global pandemic through multi-faceted measurements.
We will continue our measurements to observe the effect on phishing and scams once the crisis subsides.
To enable early characterization and detection of emerging types of phishing attacks, future research should also focus on developing real-time monitoring approaches to reliably conduct comprehensive and holistic measurements of phishing in an automated way.

%% file: relatedwork.tex
\section{Related work}
\label{s:relwk}

\subsection{Mitigating Phishing}
Phishing attacks, the most prevalent web-based threat, have caused substantial damage to victims~\cite{van2019cognitive, ho2019detecting}.
To detect and mitigate phishing attacks, much research effort focused on analyzing phishing URLs~\cite{bin2010dns, blum2010lexical, huang2012svm, khonji2011novel} and website content~\cite{wu2006web, zhang2007cantina, zhang2011textual, bilge2011exposure, canali2013role}.

Sahingoz et al.~\cite{sahingoz2019machine} proposed a method to detect phishing websites based on the URL. They extract Natural Language Processing (NLP) based features such as word counts, word length, and TLD to train a random forest classifier capable of detecting phishing URLs. While they show their proposed method outperforms previous models, adversaries can bypass URL classification algorithms~\cite{aleroud2020bypassing}. 

As content-based approaches are proved to have a better performance than URL based methods~\cite{oest2018inside}, most of the new methods focus on analyzing the page content and search engine metadata~\cite{xiang2011cantina+}. Ardi et al. proposed a content-based method for detecting phishing websites on demand. Their method leverages the Document Object Model (DOM) of a webpage to detect phish. This method breaks the DOM tree into chunks and computes the hash of each chunk. If the number of chunks that matches the hashed blacklist content is greater than a threshold, it flags the webpage as phishing. This method provides good performance and zero false positive rate~\cite{ardi2016auntietuna}. However, an attacker can simply use homographs (look-alike characters) or replace the content with images to bypass the detection method~\cite{ardi2016auntietuna}.

Google Safe Browsing~\cite{whittaker2010large} and Microsoft SmartScreen~\cite{smartscreen} are currently deployed mitigation systems across major web browsers for protecting users from phishing attacks.
They detect a phishing website based on a URL blacklist or a heuristic classifier.
As the only mitigation against phishing attacks is the blacklists, if blacklists do not offer adequate protection, users will be exposed to phishing threats without any protection~\cite{nsslabs}.

\subsection{Limitations of Current Anti-phishing Systems}
Several research works revealed the limitations of blacklist-based anti-phishing approaches~\cite{han:phisheye, oest2018inside, oest2019phishfarm, peng2019opening}.
Han et al.~\cite{han:phisheye} monitored the lifecycle of phishing websites from the creation of them by using a honeypot web server.
Oest et al.~\cite{oest2019phishfarm} conducted an empirical study on the blacklisting coverage and response time. In this work they propose the PhishFarm framework. PhishFarm first deploys different phishing websites, then it reports the deployed websites and waits for the anti-phishing entities to blacklist the reported websites. By using this framework, they test the resilience of anti-phishing entities.
Peng et al.~\cite{peng2019opening} measured the performance of the VirusTotal and its third-party vendors with their own phishing sites. They use a similar method as PhishFarm to study the reliance and robustness of VirusTotal and its 68 third-party vendors. In this work, they set up their own phishing websites while monitoring the incoming traffic and the VirusTotal labeling process.
All of them suggested a significantly faster blacklist response time for protecting users more effectively.
In addition, cybercriminals continue to use evasion techniques to make phishing websites remain online so that it is accessible to victim users for a long time~\cite{oest2018inside, oest2020sunrise, zhang2021crawlphish}.
These studies imply that, as far as the standard anti-phishing defense is operated in a reactive manner, phishing attacks will still remain a significant threat to Internet users.  

%% file: conclusion.tex
\section{Conclusion}
\label{s:conclusion}


%
Amid widespread panic and uncertainty, the increased usage of online services during the COVID-19 pandemic resulted in an early spike of online social engineering attacks.
To gain insight into how the pandemic changed trends in phishing and scams and how attackers took advantage of this situation, we synthesized multiple sources of web-related data. 
Our analysis revealed the potential for new ecosystem defenses and enhanced collaboration among entities to support a more timely and effective ecosystem strategy to combat surges in phishing volume and sudden shifts in the nature of attacks.